\documentstyle[12pt,epsfig]{article}
\newcommand{\be}{\begin{equation}}
\newcommand{\ee}{\end{equation}}
\newcommand{\bea}{\begin{eqnarray}}
\newcommand{\eea}{\end{eqnarray}}
\newcommand{\Tr}{\mathop{\rm Tr}\nolimits}
\newcommand{\arcsinh}{\mathop{\rm arcsinh}\nolimits}
\newcommand{\Str}{\mathop{\rm Str}\nolimits}

\newcommand{\xy}{x^2+y^2}
\newcommand{\xyz}{\sqrt{(x^2-y^2)^2 + z^4}}
\begin{document}
\begin{titlepage}
\begin{flushleft}
       \hfill                      {\tt hep-th/9804018}\\
       \hfill                       SISSA ref. 34/98/EP\\
\end{flushleft}
\vspace*{3mm}
\begin{center}
{\LARGE Taming the Non Abelian Born-Infeld Action\\ }
\vspace*{12mm}
{\large Irina Ya. Aref'eva }\\
\vspace{2mm}
Steklov Mathematical Institute, Russian Academy of Sciences\\
Gubkin St.8, GSP-1, 117966, Moscow, Russia\\
{\tt arefeva@mi.ras.ru},\\
\vspace{4mm}
{\large Gabriele Ferretti}\\
\vspace{2mm}
SISSA, Via Beirut 2 Trieste 34013, Italy\\
{\tt ferretti@sissa.it}\\
\vspace{4mm}
and\\
\vspace{4mm}
{\large  Alexey S. Koshelev}\\
\vspace{2mm}
Physical Department, Moscow State University, \\
Moscow, Russia, 119899 \\
{\tt kas@depni.npi.msu.su}
\vspace*{10mm}
\end{center}

\begin{abstract}

We show how to reduce the non abelian Born-Infeld action describing
the interaction of two D-particles to the sum of elliptic integrals
depending on simple kinematic invariants. This representation
gives explicitly all $\alpha^\prime$  corrections to D-particle dynamics. 
The $\alpha^\prime$ corrections induce a stabilization  of the
classical trajectories such as the ``eikonal''
which are unstable within the Yang-Mills approximation.

\end{abstract}
\end{titlepage}
\section{Introduction}

Since Dirichlet-branes (D-branes) found their true place in string
theory~\cite{P1}, they have been challenging our basic intuitions about
space-time. One of the most intriguing features is the fact that the
coordinates describing the relative positions of D-branes naturally appear
to be matrix valued~\cite{W1}, a fact that is at the heart
of the matrix theory conjecture~\cite{BFSS} (for a review see~\cite{M-rev}).

In the matrix theory limit~\cite{sen,sei}, the form of the action relevant for
the dynamics is simply the dimensionally reduced maximally supersymmetric
Yang--Mills theory~\cite{CH}-~\cite{ DKPS}
but this is not so in other regimes. A specific example
is ordinary weakly coupled type II (A or B) string theory at fixed
value of $\alpha^\prime$,
where the relevant effective action is the disk generating functional, already
a highly non local object that is not known in closed form. If one restricts
oneself further to considering the limit where the accelerations and higher
time derivatives are small (keeping however the possibility of having
relativistic velocities), the relevant action is the non abelian generalization
of the Born--Infeld action (NBI)~\cite{Ts1} to be described
below\footnote{For earlier references, see~\cite{NS}-~\cite{Ts3}.}.
The full supersymmetric extension of
such action is still out of reach, but for our purposes
the bosonic action of~\cite{Ts1} will suffice.

One of the hurdles in using the usual form of the NBI lagrangian is the
presence of the symmetrized trace operator in front of the usual square
root of the determinant. This is particularly disconcerting in the case of
a D-particle, where one would like to use this lagrangian to investigate
corrections to the scattering processes of~\cite{DKPS} or to the canonical
quantization of~\cite{DFS, KP}.

In this paper we solve this problem by showing how, in the case of two
D-particles, the NBI lagrangian can be recast in the form of an ordinary
function of a small number of kinematic invariants. The functional dependence
of the lagrangian on these invariants in through elliptic integrals.

As a first application of our result we investigate the stability of the
eikonal trajectory, which is known to be unstable in the Yang--Mills
system which is chaotic~\cite{aretal}.
We show how the $\alpha^\prime$ corrections
change the problem from the usual one of particles in flat space subjected to
a quartic potential with flat directions to that of particles in
\emph{conformally} flat space with a similar potential. The curvature
computed from the conformal factor is positive in the region near the flat
directions thus contributing to the stabilization.

Two further issues that we leave for future work are the possibility of
existence of trajectories other than the eikonal but with the same asymptotic
behavior as $|t|\to\infty$. It would be interesting to find such
trajectories where two D-particles come close, exchange their identity
through a rotation in ``non commutative space'' and then separate again.
Such trajectories do not exist in the truncated Yang--Mills theory~\cite{AFK}
but may be present for the full NBI action. Another interesting issue would
be to study the effects that the $\alpha^\prime$ corrections have on the
Born--Oppenheimer approximation, perhaps suggesting a change of variables
that makes it applicable for small distances.

The paper is organized as follows. In the next section we state
Tseytlin's result~\cite{Ts1} on the form of the NBI action that is
the starting point of our investigation. We also summarize
the basic elementary features of the symmetric trace necessary to give
a precise meaning to Tseytlin's action. In section three we show
how to rewrite the NBI action as a simple function of a few invariants
in the case of two D-particles.
We begin by stating the result and all the assumptions that go into it and
end with the actual proof. The last section contains a first look at the
dynamics that can be obtained from such an action (e.g. the stabilization
of the eikonal) and some possible future directions to be explored.

\section{The Non Abelian Born Infeld Action}

Already at tree level in the string coupling $g_s$ (disk diagram) and in flat
space-time $g_{\mu\nu}=\eta_{\mu\nu}={\rm diag}(-1,1,\cdots,1)$, the effective
action for $N$ Dp-branes is a highly complicated object that is not known in
closed form. This is of course due to the fact that, even at tree level,
the n-point function on the disk receives contributions from the massive
string states that, when integrated out, yield a non local functional
of the massless modes.

If one makes the further approximation of neglecting higher derivatives of
the field strength $F$ and, by consistency, terms involving the commutator
of two field strengths as $[F,F]\equiv [D,D]F$, one can cast the bosonic part
of the remaining action in the non abelian generalization of the Born Infeld
action (NBI)\footnote{In all other equations we shall drop the dependence
on $g_s$ and $2\pi\alpha^\prime$}:
\be
   S= -\frac{1}{2g_s}\int dx^{p+1} \Str\sqrt{-\det\bigg(\eta_{\mu\nu}
      + 2\pi\alpha^\prime F_{\mu\nu}\bigg)}, \label{NBI}
\ee
where $\mu, \nu = 0,\cdots,9$ but the field strength $F_{\mu\nu}$ only
depends on the first $p+1$ coordinates (coordinates on the Dp-brane). The
determinant $\det$ is taken only on the Lorentz indices
$\mu$ and $\nu$ and the symmetric trace $\Str$ will be discussed in
more details below.  Note that the
determinant of a matrix with non commuting elements can be defined in many
inequivalent ways; the symmetric trace $\Str$ picks out the definition of
physical interest.

\subsection{The Symmetric Trace}

The symmetric trace of a set of $n$ matrices is defined as the sum of the
ordinary traces over all possible permutations $\pi\in S^n$ with the
appropriate weight $1/n!$
\be
   \Str(A_1,\cdots,A_n)=\frac{1}{n!}\sum_{\pi\in S^n}
   \Tr(A_{\pi(1)}\cdots A_{\pi(n)}). \label{defstr}
\ee
The trivial technical point that must be understood about
the symmetric trace $\Str$
is that, contrary to the more familiar trace operator, it does not allow
one to perform the matrix algebra inside it. To give an explicit example,
let $A$, $B$ and $C$ be three arbitrary matrices
and let $D=AB$. Then, while obviously $\Tr(ABC)=\Tr(DC)$,
one has from (\ref{defstr})
\be
  \Str(A,B,C)=\frac{1}{2}\Tr(ABC+ACB)\not=\Str(D,C)=\Tr(ABC).
\ee
That's why, to avoid
any confusion it is better to write $\Str(A,B,C)$  instead of $\Str(ABC)$.
For an arbitrary function $f(A_1,\cdots,A_n)$ of $n$ matrices $A_i$, the
symmetric trace (with respect to \emph{those} matrices) is defined
by first expanding $f$ as a Taylor series and then performing the symmetric
trace on each monomial:
\be
   \Str(f(A_1,\cdots,A_n))=\sum_{k_1\cdots k_n}
   \frac{1}{k_1!\cdots k_n!}f^{(k_1,\cdots,k_n)}(0,\cdots,0)
   \Str(A_1^{(k_1)},\cdots, A_n^{(k_n)}). \label{strdef}
\ee
It is important to understand that the superscripts $(k_i)$ inside the
$\Str$ operator simply mean that the matrix $A_i$ is repeated $k_i$ times
in the list and do not represent a multiplication.

In the case of (\ref{NBI}) by $\Str$ we always refer to the symmetrized
trace with respect to the form $F$.

\section{NBI action of D-particles}

In this section we derive the explicit expression (i.e. we perform the
symmetric trace) for the NBI action
of two D-particles in a particular gauge that although simplifies some
of the computation, does not make us loose generality. Since the
computations involved are rather laborious, we will begin by explaining
the choice of gauge and stating the
result; the remainder of the section deals only with the derivation
of our expression for the NBI action.

\subsection{Choice of gauge and statement of the result}

Let us consider two D-particles moving in the
$\hat 1,\; \hat 2$ plane and let us
set $A_0=A$, $A_k=X_k$ ($k=1,2$ only) so that
\be
    F_{0k}=\frac{d}{dt} X_k - i[A,X_k]\equiv DX_k \quad\hbox{and}\quad
    F_{12}=-i[X_1,X_2]
\ee
where all fields are $SU(2)$ valued.
While keeping track of the non abelian nature of the coordinates,
one can still partially fix the gauge to
be\footnote{Upper indexes always refer to Lie algebra components, lower
indexes always refer to Lorenz components.}
\be
   X_k=x_k^1\sigma^1 + x_k^2\sigma^2,
   \quad\hbox{and}\quad A= a\sigma^3.
\ee

This gauge can be reached as follows: at each time $t$ one chooses the gauge
transformation $g(t)$ that sets the commutator $[X_1, X_2]$
proportional to $\sigma^3$. The remaining $U(1)$ gauge freedom
of rotating along the
$\sigma^1, \sigma^2$ plane is compensated by the gauge field $A=a\sigma^3$.
If the commutator is zero, the transformation above is ill defined, but in
this case $X_1$ and $X_2$ are parallel in the $SU(2)$ Lie algebra and can both
be made independent on $\sigma^3$ by a gauge rotation.

The main statement is that the NBI action in this gauge can be written in
terms of the following three quantities, invariant in the unbroken
gauge\footnote{For convenience we indicate the \emph{covariant} time
derivative of the components by a dot: $DX_k=\dot x_k^1\sigma^1 +
\dot x_k^2\sigma^2$.}
\bea
    \rho &=& 4(x_1^1 x_2^2 - x_2^1 x_1^2)^2 \nonumber\\
    \eta &=& \frac{1}{2}\Big((\dot x_1^1)^2 + (\dot x_2^1)^2 +
             (\dot x_1^2)^2 + (\dot x_2^2)^2\Big) \nonumber\\
        && -\frac{1}{2}\sqrt{\Big((\dot x_1^1)^2 + (\dot x_2^1)^2 -
             (\dot x_1^2)^2 - (\dot x_2^2)^2 \Big)^2 +
         4\Big(\dot x_1^1 \dot x_1^2 + \dot x_2^1 \dot x_2^2 \Big)^2}
          \nonumber\\
    \nu &=& \frac{1}{2}\Big((\dot x_1^1)^2 + (\dot x_2^1)^2 +
             (\dot x_1^2)^2 + (\dot x_2^2)^2\Big) \nonumber\\
        && +\frac{1}{2}\sqrt{\Big((\dot x_1^1)^2 + (\dot x_2^1)^2 -
             (\dot x_1^2)^2 - (\dot x_2^2)^2 \Big)^2 +
         4\Big(\dot x_1^1 \dot x_1^2 + \dot x_2^1 \dot x_2^2 \Big)^2}
    \label{invariants}
\eea
as
\bea
   S(\rho,\eta,\nu)&=&-\sqrt{(1+\rho)\frac{1-\nu}{1-\eta}}-
      \frac{\nu}{\sqrt{\nu+\rho}}\Big(F(\phi,m)-F(\pi/2,m)\Big)
          \nonumber \\
      &&+\sqrt{\nu+\rho}\Big(E(\phi,m)-E(\pi/2,m)\Big)
      \label{explicit}
\eea
where $F$ and $E$ are the elliptic integrals of first and second kind
respectively whose arguments are:
\be
   \phi=\arcsin\sqrt{\frac{1-\nu}{1-\eta}}\quad\hbox{and}\quad
   m=\frac{\eta+\rho}{\nu+\rho}.
\ee
Since there are different conventions in the literature, let
us also recall the definition of the elliptic integrals we are using:
\be
   F(\phi,m)=\int_0^\phi\;\frac{d\theta}{\sqrt{1-m\sin^2\theta}}
             \quad\hbox{and}\quad
   E(\phi,m)=\int_0^\phi\sqrt{1-m\sin^2\theta}\;d\theta. \label{elliptic}
\ee
$F(\pi/2,m)$ and $E(\pi/2,m)$ are known as complete elliptic integrals
and are sometimes indicated by $K(m)$ and $E(m)$.

For completeness, let us write the action of $R$, the generator of rotation
in the $\hat 1,\; \hat 2$ plane, and $G$, the generator of the left over
$U(1)$ gauge invariance.
\be
    R:\left\{\begin{array}{c}
             \pmatrix{x_1^1\cr
                     x_2^1\cr}\rightarrow \pmatrix{\tilde x_1^1\cr
                                                  \tilde x_2^1\cr}=
      \pmatrix{\cos\theta x_1^1 - \sin\theta x_2^1\cr
              \sin\theta x_1^1 + \cos\theta x_2^1\cr} \nonumber \\
                     \nonumber \\
             \pmatrix{x_1^2\cr
                     x_2^2\cr}\rightarrow \pmatrix{\tilde x_1^2\cr
                                                  \tilde x_2^2\cr}=
      \pmatrix{\cos\theta x_1^2 - \sin\theta x_2^2\cr
              \sin\theta x_1^2 + \cos\theta x_2^2\cr}
              \end{array}\right.
\ee
\be
    G:\left\{\begin{array}{c}
             \pmatrix{x_1^1\cr
                     x_1^2\cr}\rightarrow \pmatrix{\hat x_1^1\cr
                                                  \hat x_1^2\cr}=
      \pmatrix{\cos\alpha(t) x_1^1 - \sin\alpha(t) x_1^2\cr
              \sin\alpha(t) x_1^1 + \cos\alpha(t) x_1^2\cr} \nonumber \\
                      \nonumber \\
             \pmatrix{x_2^1\cr
                     x_2^2\cr}\rightarrow \pmatrix{\hat x_2^1\cr
                                                  \hat x_2^2\cr}=
      \pmatrix{\cos\alpha(t) x_2^1 - \sin\alpha(t) x_2^2\cr
              \sin\alpha(t) x_2^1 + \cos\alpha(t) x_2^2\cr} \nonumber \\
               \nonumber \\
      a\rightarrow \hat a = a - \frac{1}{2}\frac{d\alpha}{dt}
            \end{array}\right.
\ee
It is an easy matter to check that (\ref{invariants}) are invariant under
$R$ and $G$.

\subsection{Derivation of the form of the action}

The remainder if this section contains the derivation of (\ref{explicit}).

We begin by
evaluating the determinant in (\ref{NBI})
without performing the matrix algebra:\footnote{The meaning of
symbols like $(DX_1,-DX2,-i[X_1,X_2])$
is simply that of an \emph{unordered list} of matrices, just as in
(\ref{strdef}).}
\bea
   &&\det(\eta_{\mu\nu}+F_{\mu\nu})=
     \det\pmatrix{-1&DX_1&DX_2\cr
                -DX_1&1&-i[X_1,X_2]\cr
                -DX_2&i[X_1,X_2]&1\cr}\nonumber\\
   && = -1 +(DX_1,-DX_2,-i[X_1,X_2]) +(-DX_1,DX_2,i[X_1,X_2]) \nonumber\\
   &&-(1,DX_2,-DX_2) -(1,DX_1,-DX_1)
                -(-1,i[X_1,X_2], -i[X_1,X_2])\nonumber\\
   &&= -1 + (DX_k,DX_k) + ([X_1,X_2],[X_1,X_2])
\eea
where the last simplification is allowed because all these terms will be
used inside the symmetric trace.
By expanding the square root as a double power series one gets
\bea
   S&=&-\frac{1}{2}\Str\sqrt{-\det(\eta_{\mu\nu} + F_{\mu\nu})}\nonumber\\
   &=&-\frac{1}{2}\Str\sqrt{1 - (DX_k,DX_k) - ([X_1,X_2],[X_1,X_2])}
                \nonumber\\
   &=& \sum_{m,q=0}^\infty \frac{1}{2^{2m+2q}}\frac{(2m+2q-2)!}
   {m!q!(m+q-1)!}\Str(DX_k^{(2m)},[X_1,X_2]^{(2q)}),\nonumber\\
    \label{NBIpart}
\eea
where, for $m=q=0$ we define the coefficient of the symmetric trace to be
equal to its ``analytically continued'' value $-1/2$.

Eq. (\ref{NBIpart}) can be simplified by noticing that the commutator
$[X_1,X_2]$ anti-commutes with the covariant derivatives and has square
proportional to the identity matrix. This allows one to write
\be
   \Str(DX_i^{(2m)},[X_1,X_2]^{(2q)}) =
   \frac{2m!2q!}{(2q+2m)!}\frac{(m+q)!}{m!q!}(-1)^q \rho^q\Str(DX_i^{(2m)})
   \label{firstsimpli}
\ee
in terms of the invariant $\rho$ defined in (\ref{invariants}).
Note that we are still using $\Str$.

The reader can check the validity of (\ref{firstsimpli}) by chosing the term
in the sum containing only $DX_1$, expanding the symmetric trace
into $(2m+2q)!$ ordinary traces according
to (\ref{defstr}) and grouping the $2m!2q!$ identical terms coming
from the permutations of $DX_1$ with themselves and $[X_1, X_2]$ with
themselves. In each of the remaining $(2m+2q)!/2m!2q!$ terms one then brings
all the commutators $[X_1,X_2]$ to the left keeping track of the signs.
There is an excess of $(m+q)!/m!q!$ terms with positive sign, yielding
\be
   \Str(DX_1^{(2m)},[X_1,X_2]^{(2q)}) =
   \frac{2m!2q!}{(2q+2m)!}\frac{(m+q)!}{m!q!}(-1)^q \rho^q\Tr(DX_1^{2m}).
     \label{tem}
\ee
Eq. (\ref{firstsimpli}) follows by replacing $(DX_1)^{(2)}$ by
$(DX_i)^{(2)}$ and $\Tr$ by $\Str$ on the r.h.s. of (\ref{tem}).

With this simplification:
\be
   S = \sum_{m,q=0}^\infty \frac{1}{2^{2m+2q+1}}\frac{2m! 2q!}
   {m!^2 q!^2}
   \frac{(-1)^q}{2q+2m-1}\rho^q\Str(DX_i^{(2m)}). \label{partialsum}
\ee

To proceed further, define three (non invariant) intermediate quantities
\bea
    && a = (\dot x_1^1)^2 +  (\dot x_2^1)^2 \nonumber\\
    && b = 2(\dot x_1^1 \dot x_1^2 + \dot x_2^1 \dot x_2^2) \nonumber\\
    && c = (\dot x_1^2)^2 +  (\dot x_2^2)^2
      \label{abc}
\eea
and carefully expand the symmetric trace using (\ref{defstr})
\bea
    &&\Str(DX_i^{(2m)})=\Str\bigg(\Big((\dot x_1^1\sigma^1+
      \dot x_1^2\sigma^2)^{(2)}+(\dot x_2^1\sigma^1+
      \dot x_2^2\sigma^2)^{(2)} \Big)^{(m)} \bigg)\nonumber\\
    &&=\Str\bigg(\Big(a(\sigma^1,\sigma^1) +
                     b(\sigma^1,\sigma^2) +
                     c(\sigma^2,\sigma^2)\Big)^{(m)}\bigg)\nonumber\\
    &&=\sum_{i+j+k=m}\frac{m!}{i!j!k!}a^i b^j c^k
      \Str\Big((\sigma^1)^{(2i+j)}(\sigma^2)^{(2k+j)} \Big)\nonumber\\
    &&=\sum_{i+j+k=m}2\frac{m!^2}{2m!}\frac{a^i b^j c^k}{i!j!k!}
      \left(\frac{1+(-1)^j}{2}\right)\frac{(2i+j)!(2k+j)!}
      {(i+j/2)!(k+j/2)!}. \label{secondsimpli}
\eea
The reason for the combinatorics in the last line of
(\ref{secondsimpli}) is exactly the
same as the one given for (\ref{firstsimpli}); also note that the symmetric
trace vanishes for $j$ odd explaining the presence of the projection
operator $(1+(-1)^j)/2$ and term of the type $j/2$ in the factorial.
In fact, by using the identity
\be
   \sum_{m=0}^\infty\sum_{i+j+k=m} a_{m,i,j,k} =
   \sum_{i,j,m =0}^\infty a_{m+i+j,i,j,m}
\ee
and replacing $j$ by $2j$, we have the following multiple power series
for the NBI action:
\be
   S=\sum_{q,m,i,j=0}^\infty \frac{(-1)^q}{2^{2m+2i+4j+2q}}
   \frac{2q!}{q!^2}
   \frac{(2i+2j)!(2m+2j)!}{i!2j!m!(i+j)!(m+j)!}
   \frac{\rho^q a^ib^{2j}c^m}{2m+2i+4j+2q-1}.
\ee
To simplify things slightly, let us introduce four new variables $x,y,z,w$
and define
\be
   U(x,y,z,w)=\sum_{q,m,i,j=0}^\infty
   \frac{2q!}{q!^2}
   \frac{(2i+2j)!(2m+2j)!}{i!2j!m!(i+j)!(m+j)!}
   \frac{x^{2m} y^{2i} z^{4j} w^{2q}}{2m+2i+4j+2q-1}
   \label{powerU}
\ee
Trivially, if one knows $U$, the NBI action is
\be
   S = U(\sqrt{c}/2, \sqrt{a}/2, \sqrt{b}/2, i\sqrt{\rho}/2).
    \label{SU}
\ee

To obtain a closed expression for $U$ notice that there is a very similar
expression that can be easily summed:
\bea
   &&Z(x,y,z,w)=\sum_{q,m,i,j=0}^\infty
   \frac{2q!}{q!^2}
   \frac{(2i+2j)!(2m+2j)!}{i!2j!m!(i+j)!(m+j)!}
   x^{2m} y^{2i} z^{4j} w^{2q}\nonumber\\
   &&=\frac{1}{\sqrt{(1-4w^2)\Big((1-4x^2)(1-4y^2)-4z^4\Big)}}.
\eea

The functions $U$ and $Z$ are related by the PDE:
\be
   \Big(x\partial_x+y\partial_y+z\partial_z+w\partial_w-1\Big)U = Z
   \label{PDE}
\ee
and we are left with the problem of finding the solution to (\ref{PDE})
compatible with the conditions given by the power series (\ref{powerU}).
Let us first find the most general solution to (\ref{PDE}) by making the
change of variable
\be
   p=(xyzw)^{1/4}, \quad r=\left(\frac{w}{x}\right)^{1/4},
     \quad q=\left(\frac{w}{y}\right)^{1/4}
     \quad s=\left(\frac{w}{z}\right)^{1/4}
\ee
that reduces the PDE to an ODE
\be
   \Big(p\partial_p -1\Big) U =  Z
   \label{ODE}
\ee
whose most general solution in terms of an arbitrary function $g(r,q,s)$
is
\be
   U = p\left(g(r,q,s)+\int^p\frac{Z}{p^2}\;dp\right).
\ee

Performing the elliptic integral we obtain\footnote{the definition of the
elliptic integrals is given in (\ref{elliptic})}
\bea
    U &=& pg(r,q,s) - \sqrt{\frac{1-2\Big(\xy +\xyz\Big)}
         {1-2\Big(\xy-\xyz\Big)}(1-4w^4)}\nonumber\\
    &&-\frac{2\Big(\xy+\xyz\Big)}{\sqrt{2\Big(\xy+\xyz\Big)-4w^2}}\;F(\phi,m)
           \nonumber\\
    &&+\sqrt{2\Big(\xy+\xyz\Big) - 4w^2}\;E(\phi,m)
\eea
with
\bea
    \phi&=&\arcsin\sqrt{\frac{1-2\Big(\xy+\xyz\Big)}
         {1-2\Big(\xy-\xyz\Big)}}\nonumber\\
     m &=& \frac{2\Big(\xy-\xyz\Big)-4w^2}{2\Big(\xy+\xyz\Big)-4w^2}.
\eea

The function $g(r,q,s)$ can be fixed by imposing that the solution should not
have any term linear in $x,y,z,w$ in its power series around the origin.
Hence, by computing $U$ along the line $x=t/r^4,\;y=t/q^4,\;z=t/s^4,\;w=t$
and imposing the vanishing of the $O(t)$ terms one obtains
\bea
   pg(r,q,s)&=&\frac{2\Big(\xy+\xyz\Big)}
       {\sqrt{2\Big(\xy+\xyz\Big) - 4w^2}}\;F(\pi/2,m)
    \nonumber\\
    &&-\sqrt{2\Big(\xy+\xyz\Big) - 4w^2}\;E(\pi/2,m).
\eea

Tracing back the various changes of variables to (\ref{invariants})
(c.f.r. also (\ref{abc}) and (\ref{SU}) we
obtain our final answer (\ref{explicit}). Note that the three non invariant
quantities $a$, $b$ an $c$ have combined into two invariants $\nu$ and $\eta$
($\rho$ was invariant from the beginning)
so that the final expression (\ref{explicit}) is depending
only on the three variables (\ref{invariants}) and not four.

\section{Some simple properties of the action}

To get a first feeling for the quantities $\rho$, $\eta$ and $\nu$,
note that, in the commuting limit $\rho=\eta=0$ and $\nu=v^2$, $v$
being the
relative velocity of the two D-particles. Substituting these values into
(\ref{explicit}), one finds the expected result
\be
   S = -\sqrt{1-\nu} = -\sqrt{1-v^2}.
\ee
Another trivial limit is that where all time dependence is dropped, i.e.
$\eta=\nu=0$, $\rho= -\Tr([X_1,X_2]^2 )$ where one finds
the potential
\be
   S = -\sqrt{1+\rho}= -\frac{1}{2}\Tr\sqrt{1-[X_1,X_2]^2}. \label{NBIpot}
\ee
With the proper factors of the string slope reinstated,
to leading order in $\alpha^\prime$ (\ref{NBIpot}) is simply the Yang-Mills
potential and $\rho$ measures the distance from the bottom of the valley.

A less trivial limit comes by asking what is the form of the kinetic term
exact to all orders in $\alpha^\prime$. Here one already encounters an
expression that could not have been guessed simply from (\ref{NBI}).
Although it is possible to obtain this from the full answer
(\ref{explicit}), we find it easier to go back to the partial sum
(\ref{partialsum}) and specialize to the $m=1$ case. The coefficient of
\be
   \Str(DX_i^{(2)})\equiv\Tr(DX_i^2)=
   2\Big((\dot x_1^1)^2 + (\dot x_2^1)^2 +
   (\dot x_1^2)^2 + (\dot x_2^2)^2\Big)
\ee
 is
\be
   f_1(\rho)=\sum_{q=0}^\infty \frac{1}{2^{2q+2}}\frac{2q!}
   {q!^2}\frac{(-1)^q}{2q+1}\rho^q=\frac{\arcsinh(\sqrt{\rho})}
   {4\sqrt{\rho}}.
\ee
Thus, for small velocities but to all orders in $\alpha^\prime$, the
lagrangian governing the dynamics of two D-particles in this gauge is
given by
\bea
   {\cal L}&=& \frac{\arcsinh\bigg(2(x_1^1 x_2^2 - x_2^1 x_1^2)\bigg)}
   {4(x_1^1 x_2^2 - x_2^1 x_1^2)}\bigg((\dot x_1^1)^2+(\dot x_2^1)^2+
   (\dot x_1^2)^2 + (\dot x_2^2)^2\bigg) \nonumber\\
   && -\sqrt{1+4(x_1^1 x_2^2 - x_2^1 x_1^2)^2}. \label{smallv}
\eea
Eq. (\ref{smallv}) describes the motion in a conformally flat space.

It is possible to use the same change of variables introduced in
~\cite{aretal} to eliminate two of the four variables in (\ref{smallv})
by virtue of the symmetries $R$ and $G$.
Setting
\bea
    x_1^1 &=&  f\cos\theta\cos\phi - g\sin\theta\sin\phi \nonumber\\
    x_1^2 &=& -f\cos\theta\sin\phi - g\sin\theta\cos\phi \nonumber\\
    x_2^1 &=&  f\sin\theta\cos\phi + g\cos\theta\sin\phi \nonumber\\
    x_2^2 &=& -f\sin\theta\sin\phi + g\cos\theta\cos\phi \label{changevar}
\eea
and eliminating the cyclic coordinates $\theta$ and $\phi$ through their
equations of motion\footnote{$l$ is a constant of motion corresponding 
to the orbital angular momentum.}
\bea
    (f^2 + g^2)\dot\phi + 2fg\dot\theta &=& 0 \nonumber\\
    2fg \dot\phi + (f^2+g^2) \dot\theta &=& l\frac{2fg}{\arcsinh(2fg)}
\eea
one obtains the Routhian (still denoted by ${\cal L}$)
\be
\label{Rou}
   {\cal L} = \frac{1}{2}K\left(\dot f^2 + \dot g^2 \right) - V,
\ee
where
\be
   K = \frac{\arcsinh(2fg)}{2fg}\quad\hbox{and}\quad
   V = \frac{l^2}{K^2}\frac{f^2 + g^2}{(f^2 - g^2)^2} + \sqrt{1 + 4f^2g^2}.
\ee

In these coordinates the eikonal takes the form
\be
g=0,\quad\phi=0,\quad f=\sqrt{v^2t^2 + b^2},\quad \theta=\arctan(vt/b),
\ee
where $v$ and $b$ are the relative velocity and impact parameter respectively.

It is interesting to compute the curvature associated to the conformal
factor $K$. This is partly because a previous analysis ~\cite{aretal} has
shown that the eikonal trajectories are in fact unstable
in the Yang--Mills case, the system being
chaotic. A positive curvature would have a stabilizing effect on the
classical trajectories.
The Gauss curvature coming from the conformal factor $K$ turns out to be
\be
   R = -2(f^2 + g^2)\frac{(\log K)''}{K},
\ee
and it is indeed positive near the eikonal, thus having a balancing
effect on the small fluctuations around it.

To study the balancing effect we analyze the system (\ref{Rou})
numerically using  Poincar\'e sections.  The Poincar\'e
section is a surface in phase space.
Our phase space is four dimensional but energy conservation
restricts the motion to be on a three dimensional subspace.
One can then impose any constraint and fix a two dimensional
surface named Poincar\'e section.
We use the constraint $g=0$.  A point is
plotted each time the trajectory goes through the surface.
A chaotic distribution of
points says that the system under consideration is chaotic.
If there are solid or dotted lines on the Poinacr\'e
section   the system is regular.

\begin{figure}
\epsfig{file=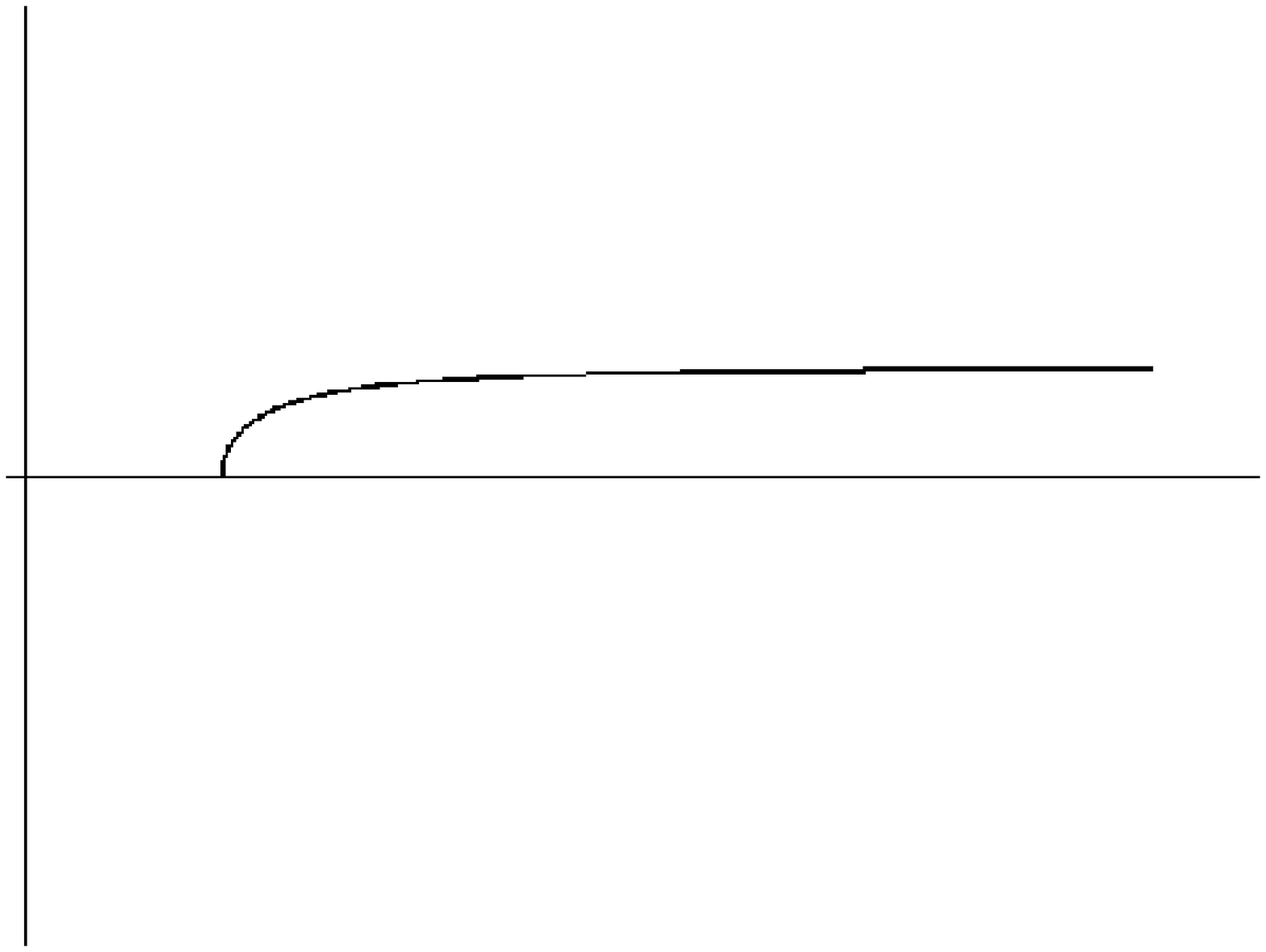,height=100mm}
\caption{Poincar\'e section for trajectory with initial data $f=5,g=0$.}
\label{fig0}
\end{figure}

\begin{figure}
\epsfig{file=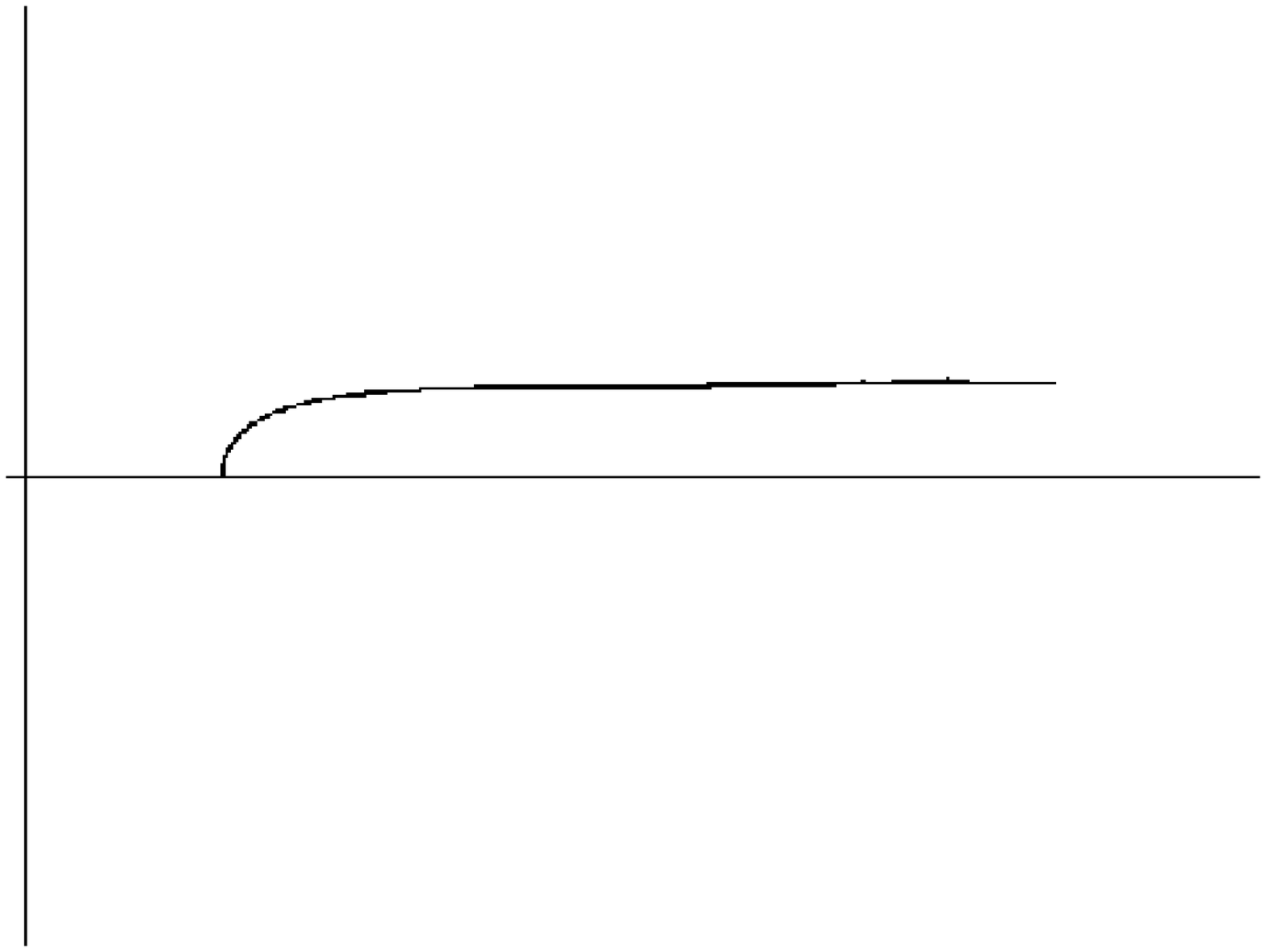,height=100mm}
\caption{Poincar\'e section for trajectory with initial data $f=5,g=0.01$.}
\label{fig01}
\end{figure}

\begin{figure}
\epsfig{file=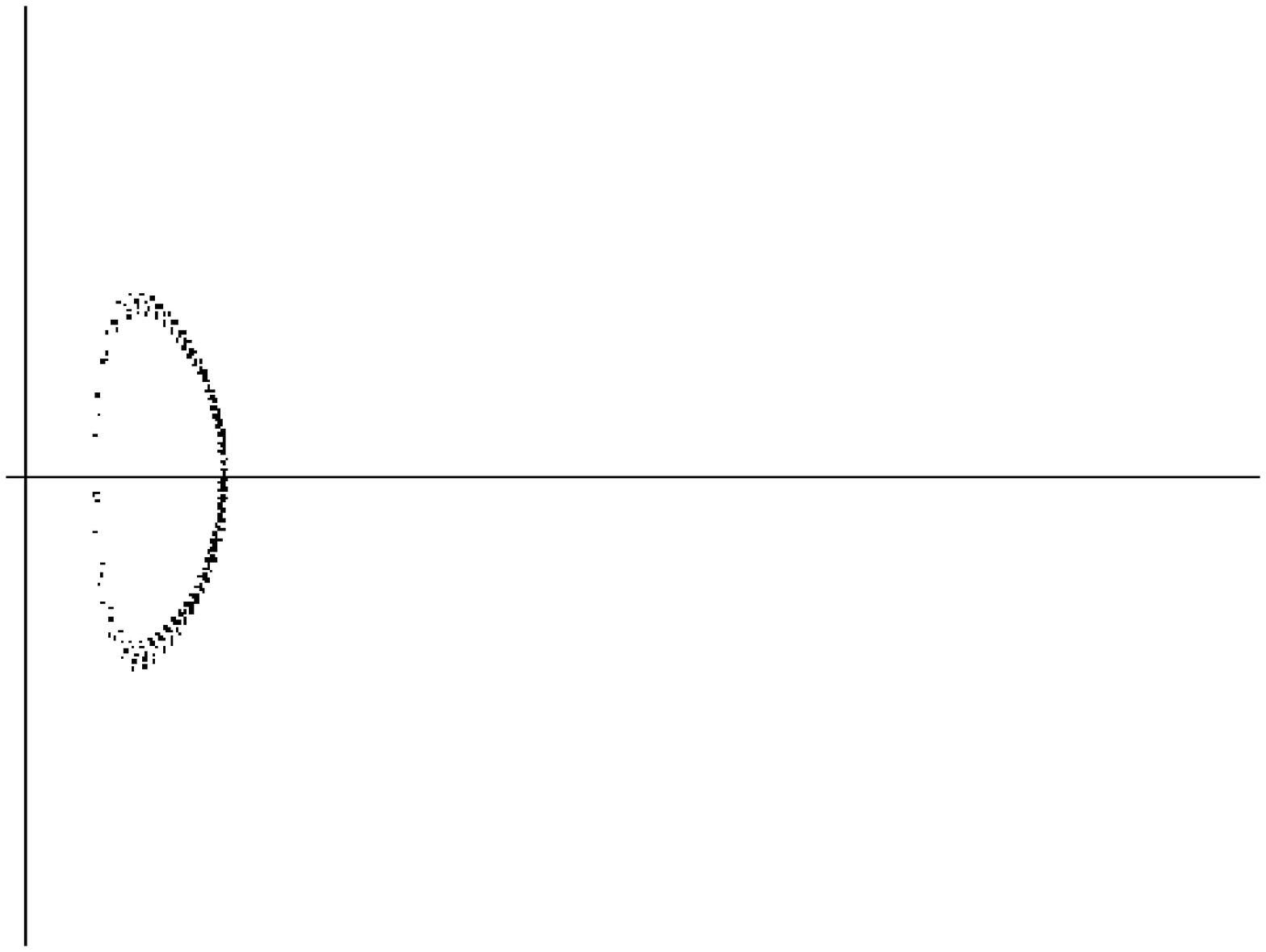,height=100mm}
\caption{Poincar\'e section for trajectory with initial data $f=5,g=0.1$.}
\label{fig1}
\end{figure}

\begin{figure}
\epsfig{file=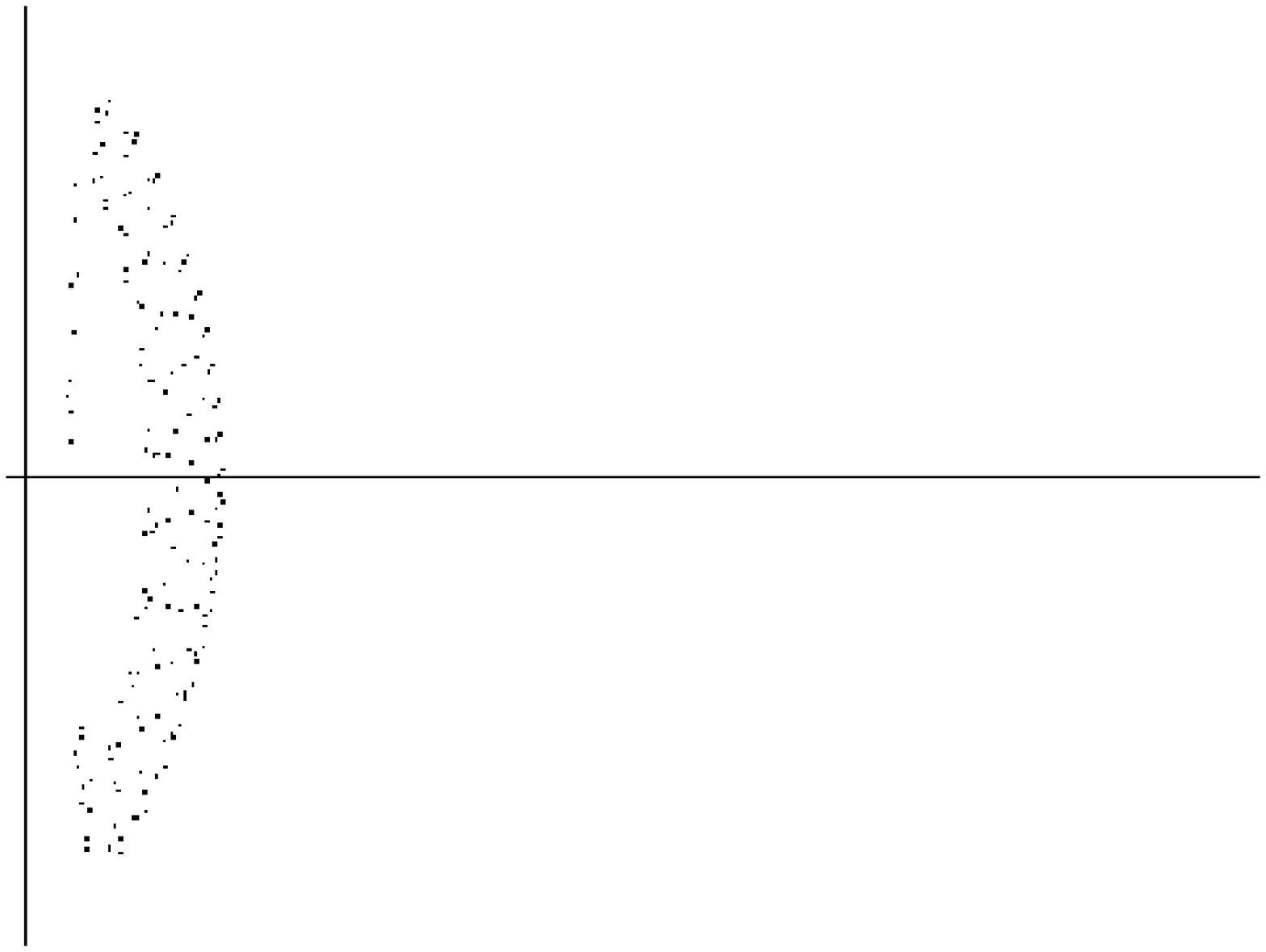,height=100mm}
\caption{Poincar\'e section for trajectory with initial data $f=5,g=0.2$.}
\label{fig2}
\end{figure}

\begin{figure}
\epsfig{file=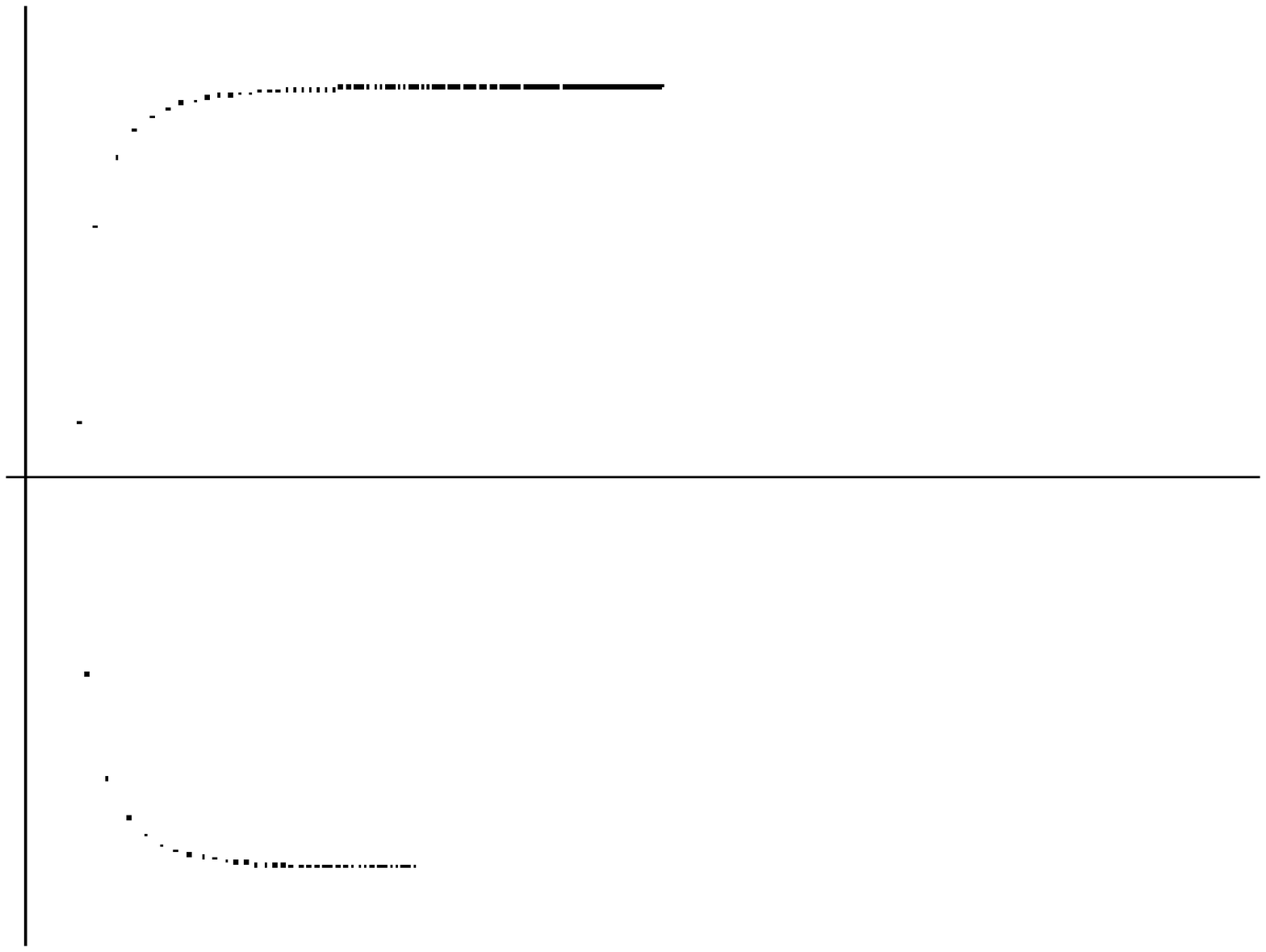,height=100mm}
\caption{Poincar\'e section for trajectory with initial 
               data $f=10,g=0.01$ and $\dot f=-1$.}
\label{fig101}
\end{figure}

\begin{figure}
\epsfig{file=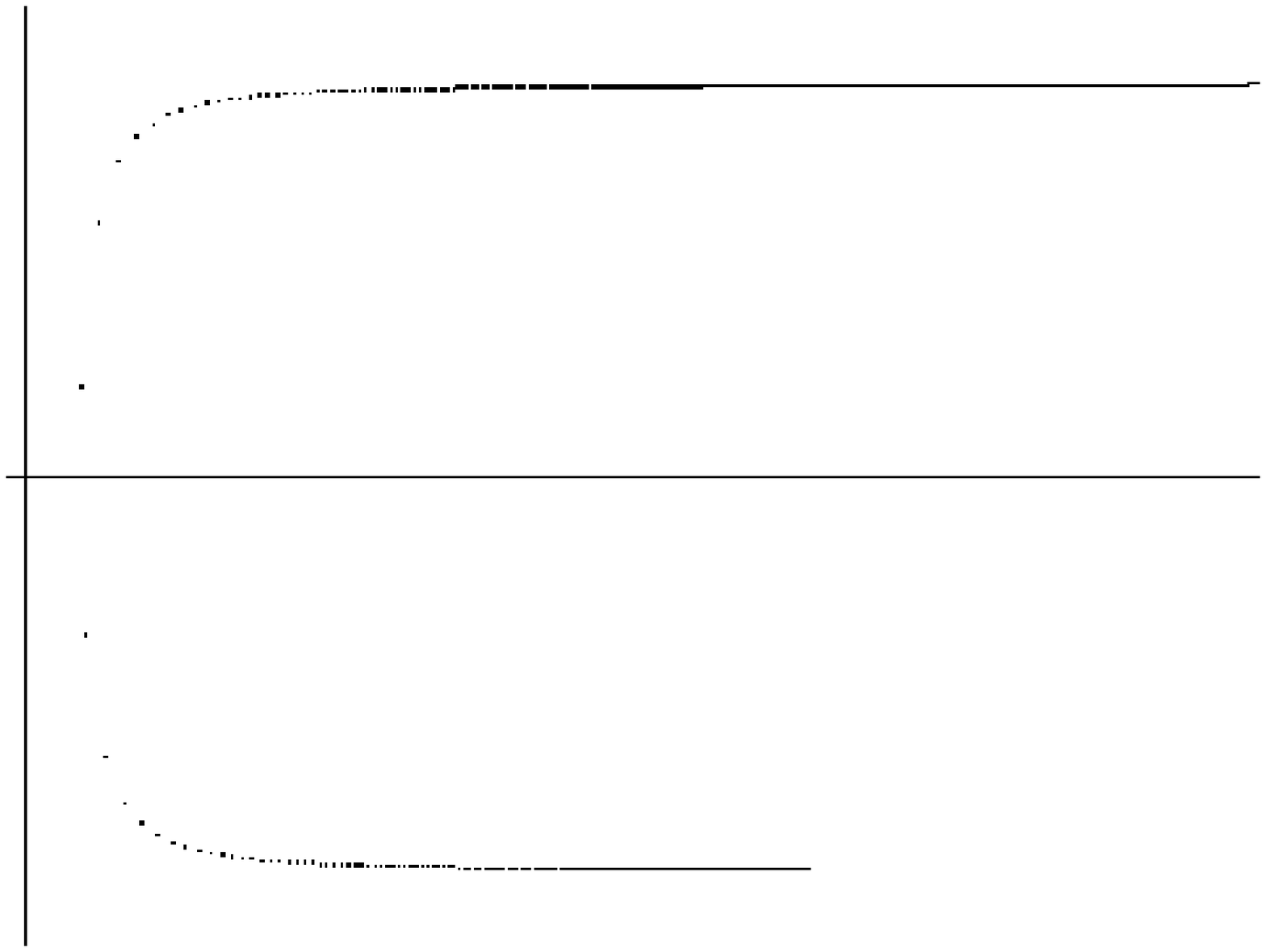,height=100mm}
\caption{Poincar\'e section for trajectory with initial 
                data $f=20,g=0.001$ and $\dot f=-1$.}
\label{fig201}
\end{figure}

We perform about 1000 experiments for 10 different values of the energy.
We plot the typical Poincar\'e sections to show an appearance  
of the regular motion.  Figures \ref{fig0} and \ref{fig01} show
the Poincar\'e
sections for  trajectories started at point $f=5,g=0$ and
$f=5$, $g=0.01$, respectively.  Comparing these
figures we see that the eikonal trajectory is stable.
Figures \ref{fig1} and \ref{fig2}
represent stochastic behavior of trajectories with the initial
coordinates $f=5$, $g=0.1$
and $f=5$, $g=0.2$, respectively.
All trajectories above have zero initial velocities.
Figures \ref{fig101} and \ref{fig201} show the motion of
the particle with initial conditions $f=10,$ $g=0.01$, $\dot f=-1$
and $f=20,$  $g=0.001$, $\dot f=-1$.
One can see that the particle reaches a point of minimal $f$,
"scatters" and turns back without stochastization.

It would be interesting to pursue the study of (\ref{explicit}) or some of
its simplified limits like (\ref{Rou}) further, 
in particular addressing some of the
issues raised in the introduction such as the detail study of trajectories
that resemble the eikonal only for $|t|\to \infty$ or ways of improving on the
Born-Oppenheimer approximation near $\rho=0$, interesting for the study of
Matrix black holes ~\cite{LM1}-\cite{LM2}.

\section{Acknowledgments}

We would like U. Danielsson, R. Iengo, J. Kalkkinen, P.B.Medvedev,
O.A. Rytchkov, A. Schwimmer, and I.V. Volovich
for discussions. I.A. and A.K. are supported
by RFFI grant 96-01-00608, I.A. is supported by INTAS grant 96-

%%pictures
\newpage
\newpage
\newpage
\end{document}